\documentclass[lettersize,journal]{IEEEtran}
\usepackage{amsmath,amsfonts}
\usepackage{algorithmic}
\usepackage{algorithm}
\usepackage{array}
\usepackage[caption=false,font=normalsize,labelfont=sf,textfont=sf]{subfig}
\usepackage{textcomp}
\usepackage{stfloats}
\usepackage{url}
\usepackage{verbatim}
\usepackage{graphicx}
\usepackage{cite}
\usepackage{booktabs} 
\usepackage{soul}
\usepackage{color, xcolor} 
\begin{document}

\title{Synesthesia of Machines Based\\ Multi-Modal Intelligent V2V Channel Model}

\author{Zengrui Han,~\IEEEmembership{Graduate Student Member,~IEEE,} Lu Bai,~\IEEEmembership{Member,~IEEE,} Ziwei Huang,~\IEEEmembership{Member,~IEEE,} and \\Xiang Cheng,~\IEEEmembership{Fellow,~IEEE} 

\thanks{Z. Han, Z. Huang, and X. Cheng are with the State Key Laboratory of Advanced Optical Communication Systems and Networks, School of Electronics, Peking University, Beijing, 100871, P. R. China (email: zengruihan@stu.pku.edu.cn, ziweihuang@pku.edu.cn, xiangcheng@pku.edu.cn).}
\thanks{L. Bai is with the Joint SDU-NTU Centre for Artificial Intelligence Research (C-FAIR), Shandong University, Jinan, 250101, P. R. China (e-mail: lubai@sdu.edu.cn).}}

% \markboth{IEEE Transactions on Wireless Communications,~Vol.~xx, No.~xx, XX~2024}
% {Submitted paper}

\maketitle

\begin{abstract}
This paper proposes a novel sixth-generation (6G) multi-modal intelligent vehicle-to-vehicle (V2V) channel model from light detection and ranging (LiDAR) point clouds based on Synesthesia of Machines (SoM).
To explore the mapping relationship between physical environment and electromagnetic space, a new V2V high-fidelity mixed sensing-communication integration simulation dataset with different vehicular traffic densities (VTDs) is constructed.
Based on the constructed dataset, a novel scatterer recognition (ScaR) algorithm utilizing neural network SegNet is developed to recognize scatterer spatial attributes from LiDAR point clouds via SoM.
In the developed ScaR algorithm, the mapping relationship between LiDAR point clouds and scatterers is explored, where the distribution of scatterers is obtained in the form of grid maps.
Furthermore, scatterers are distinguished into dynamic and static scatterers based on LiDAR point cloud features, where parameters, e.g., distance, angle, and number, related to scatterers are determined.
Through ScaR, dynamic and static scatterers change with the variation of LiDAR point clouds over time, which precisely models channel non-stationarity and consistency under different VTDs.
Some important channel statistical properties, such as time-frequency correlation function (TF-CF) and Doppler power spectral density (DPSD), are obtained.
Simulation results match well with ray-tracing (RT)-based results, thus demonstrating the necessity of exploring the mapping relationship and the utility of the proposed  model.
\end{abstract}

\begin{IEEEkeywords}
6G, Synesthesia of Machines (SoM), vehicle-to-vehicle (V2V) channel model, SegNet, scatterer recognition (ScaR).
\end{IEEEkeywords}

\section{Introduction}

\IEEEPARstart{I}{T} is well known that vehicle-to-vehicle (V2V) communications, which play an increasingly important role in intelligent transportation systems (ITSs), are anticipated to enable wider applications related to future vehicles~\cite{ITS1,ITS2}.
To enhance the safety and efficiency of ITSs, it becomes imperative for sixth-generation (6G) V2V communications to acquire a comprehensive understanding of the environment and engage in advanced V2V channel modeling~\cite{cheng2022channel}.
Existing V2V channel models can generally be classified into two main types, i.e., stochastic models and deterministic models~\cite{6Gsurvey}.
Stochastic models exploit statistical methods to randomly generate and determine channel parameters~\cite{cai2017hough,yang2019cluster}. Although stochastic models are of low complexity, the accuracy is limited due to the random generation and determination of channel parameters.
Unlike stochastic models, deterministic models aim to reproduce the procedure of physical radio propagation in the site-specific scenario. Deterministic models are of high accuracy, whereas they are computationally intensive.
As a result, stochastic and deterministic models cannot effectively achieve a decent trade-off between accuracy and complexity to support the design of 6G V2V systems in ITSs.

With the development of artificial intelligence (AI) and advanced technologies, a series of AI-based channel models for 6G V2V communications have been proposed~\cite{AImodeling}.
The powerful ability of AI to process vast data and identify complex patterns can significantly enhance the trade-off between modeling accuracy and complexity.
To support channel modeling in dynamic V2V communications, the authors in~\cite{MLcluster} proposed a method for multipath component (MPC) clustering, which conducted  MPC tracking based on the Kalman filtering and Kuhn-Munkres algorithm.
The proposed model in~\cite{MLcluster} recognized dynamic behaviors of MPCs and identified time-varying clusters.
To further obtain accurate and real-time angle-of-arrival (AoA) in high-mobility V2V channels, a fast machine learning (ML)-based AoA recognition model was proposed in~\cite{fastaoa}, including an offline training phase and an online estimation phase.
In addition to the aforementioned methods \cite{MLcluster,fastaoa}, the neural network is a powerful tool, which can provide superior capabilities in the channel modeling.
In~\cite{ANNPL}, artificial neural network (ANN) based path loss and shadowing V2V channel models were proposed, which adequately leveraged the information processing capabilities of ANNs. Furthermore, the authors in~\cite{cGANpower} proposed a V2V channel model based on generative adversarial network (GAN) and feedforward neural network (FNN), which were utilized to model the probability density function (PDF) and power of received signals. 
To further increase the modeling accuracy, the authors in \cite{GANDT} carried out the V2V channel measurement campaign and proposed a GAN-based V2V channel model based on the accurate measurement data.
Therefore, with the help of AI technologies, the accuracy of the aforementioned V2V channel models in~\cite{MLcluster,fastaoa,ANNPL,cGANpower,GANDT} can be enhanced. 
However, the aforementioned work solely utilizes uni-modal radio-frequency (RF) communication information and models are essentially black boxes, which ignore the mapping relationship between physical environment and electromagnetic space. 
This limitation exacerbates the model's excessive reliance on the dataset and the data collection scenario, resulting in low scalability. 
To overcome this limitation, a more in-depth understanding of mapping relationship between electromagnetic space and physical environment is urgently required, which can facilitate more accurate, general, and low-complexity V2V channel modeling~\cite{TVT-Huang}.

Fortunately, integrated sensing and communications (ISAC)~\cite{kumari2017ieee,li2017joint,nartasilpa2018communications}, which can provide a more in-depth insight into the electromagnetic space and physical environment, has received extensive attention in 6G V2V communications.
The integration of RF-based radar sensing and communications is feasible due to the strong similarities in hardware structure and antenna structure~\cite{paul2016survey,ISAC1}.
The emergence of ISAC can offer new opportunities for channel modeling.
Furthermore, 6G autonomous vehicles in ITSs will be equipped with multi-modal devices, including millimeter wave (mmWave) radar, light detection and ranging (LiDAR), red-green-blue-depth (RGB-D) camera, and communication device.
Compared to solely utilizing RF-based radar sensing in ISAC, non-RF sensors, e.g., LiDAR, can capture more diverse environmental features, which can be leveraged to enhance communication performance and sensing functionality.
To adequately utilize multi-modal information from communication devices and various sensors, inspired by human synesthesia, Cheng et al.~\cite{SoM} proposed a novel concept, i.e., Synesthesia of Machines (SoM), which explicitly outlined the aim of intelligent multi-modal sensing-communication integration via ANNs.

Inspired by SoM, we conduct multi-modal intelligent V2V channel modeling, which utilizes multi-modal information from communication devices and LiDAR to explore the mapping relationship between physical environment and electromagnetic space.
Based on a new mixed  sensing-communication simulation dataset, a novel scatterer recognition (ScaR) algorithm from LiDAR point clouds is developed via SoM.
The explored mapping relationship deepens the understanding of the environment and channel, which can address the aforementioned ``black box" issue and reduce the reliance on specific datasets and scenarios.
By leveraging the developed ScaR algorithm, a novel multi-modal intelligent V2V channel model is proposed, which further precisely captures channel non-stationarity and consistency. 
The major contributions and novelties of this paper are summarized as follows.

\begin{enumerate}
    \item A new V2V high-fidelity mixed sensing-communication integration simulation dataset with different VTDs is constructed, including LiDAR point clouds and channel data. 
    Based on the constructed dataset, a ScaR algorithm is developed to explore the mapping relationship between physical environment and electromagnetic space for the first time.  Furthermore, a novel time non-stationary and consistent multi-modal intelligent V2V channel model is further proposed.
    \item In the developed ScaR algorithm, scatterer spatial attributes are recognized from LiDAR point clouds via SoM~\cite{SoM}. Specifically, visibility region (VR)~\cite{huang2024lidar} and grid map are employed to reduce dimensionality and extract features of LiDAR point clouds. Furthermore, the customized network, which builds on the SegNet architecture, is adopted to explore the nonlinear mapping relationship between LiDAR point clouds and scatterers for the first time.
    \item In the proposed model based on the multi-modal information from communication devices and LiDAR, recognized scatterers are further distinguished into dynamic and static scatterers, which change with the variation of LiDAR point clouds over time. 
    Therefore, time non-stationarity and consistency are precisely modeled in consideration of vehicular traffic densities (VTDs). 
    Scatterer spatial attributes obtained from the mapping relationship overcome the limitation of the existing channel models in~\cite{jiang2023hybrid,yuan20153d,huang2024lidar,huangbuchong} that focus on modeling the statistical distribution of scatterers.
    \item Simulation results show that the developed ScaR algorithm has a grid classification accuracy of over 93\% and a prediction accuracy of over 85\% for the number of scatterers. 
    Excellent agreement between simulation results and ray-tracing (RT)-based results demonstrates the necessities of the mapping relationship exploration and the validity of the proposed model.
\end{enumerate}

The remainder of this paper is organized as follows.
Section II introduces the mixed sensing-communication integration simulation dataset.
The ScaR algorithm, which can explore mapping relationship between the physical environment and electromagnetic space, is developed in Section III.
Section IV describes the proposed multi-modal intelligent V2V channel model.
Key channel statistical properties are given in Section V. 
In Section VI, the simulation result is presented and is further compared with RT-based results.
Finally, Section VII draws the conclusion.

\section{Scenario and Dataset}
In this section, a high-fidelity mixed sensing-communication integration simulation V2V dataset is constructed based on a new communication and sensing information generation and acquisition platform~\cite{M3SC}, which contains two accurate simulation software, i.e., Wireless InSite~\cite{InSite} and AirSim~\cite{shah2018airsim}.
Wireless InSite is a platform that utilizes the RT technology to analyze radio wave propagation and wireless communication systems. 
AirSim is an open-source platform constructed on Unreal Engine and can collect high-fidelity sensing data.
The 3D coordinate of each vehicle is set snapshot by snapshot to support the continuous movement of multi-vehicles.
The size and movement trajectory of each object are detected, and thus they are the same in physical environment and electromagnetic space.

\subsection{Scenario Construction}

\begin{figure*}[t]
\centering
\includegraphics[width=7in]{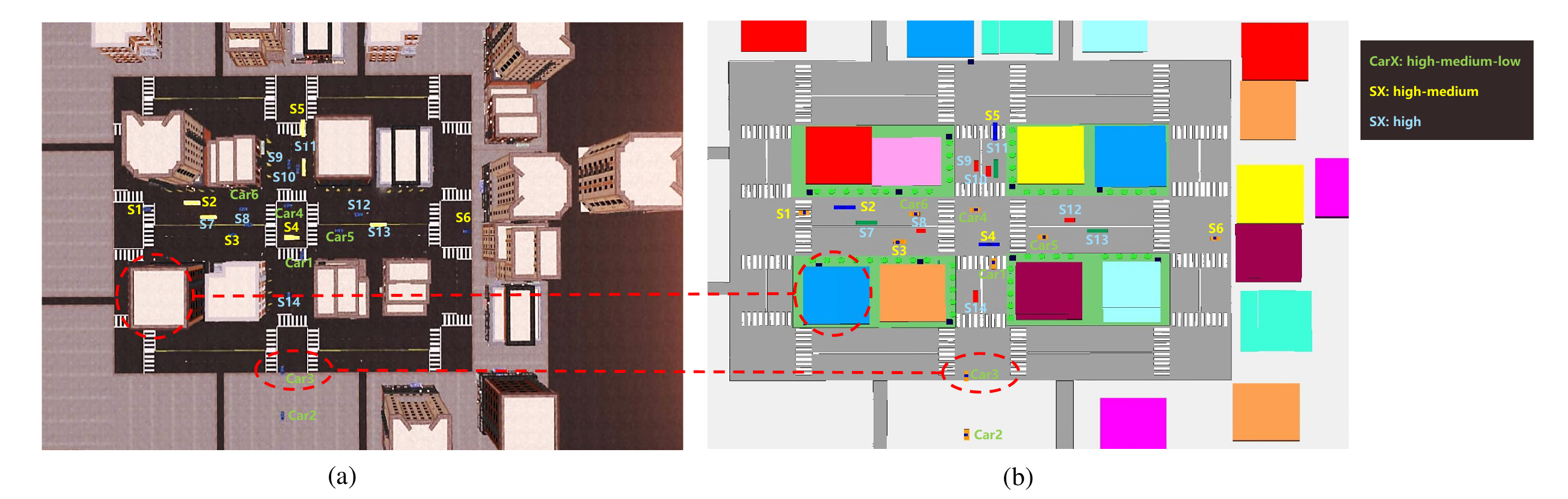}
\caption{Urban crossroads in AirSim and Wireless InSite, where vehicles marked in green appear in high, medium, and low VTDs, vehicles marked in yellow appear in high and medium VTDs, and vehicles marked in blue appear in high VTD. (a) Scenarios in AirSim under high, medium, and low VTDs.
(b) Scenarios in Wireless InSite under high, medium, and low VTDs.}
\label{WI_AirSim_scenario}
\end{figure*}

A typical vehicular scenario, i.e., urban crossroad, is constructed. The physical environment is constructed in AirSim. 
By importing 3D models from AirSim into Wireless InSite, the electromagnetic space is constructed, which precisely aligns with the physical environment to guarantee alignment between sensing data and communication data.
In Fig.~\ref{WI_AirSim_scenario}, urban crossroad scenarios at snapshot 800 under different VTDs in AirSim and Wireless InSite are given. 
In Fig.~\ref{WI_AirSim_scenario}(a), the high-fidelity physical environment in AirSim under high, medium, and low VTDs is depicted, including dynamic vehicles, static buildings, and static trees.
By simplifying the physical environment that contains detailed environmental information, the electromagnetic space in Wireless InSite is presented in Fig.~\ref{WI_AirSim_scenario}(b), thus achieving the in-depth integration of physical environment and electromagnetic space.
The urban scenario includes high/medium/low VTDs and different communication links under different street types. 
The numbers of vehicles in high, medium, and low VTDs are 20, 12, and 6, respectively. 
The types of streets include horizontal, vertical, and intersecting, which enrich the diversity of the scenario.

\subsection{Scenario Generation and Data Collection}

The number of snapshots is set to 1500 for each VTD condition. 
Coordinates of vehicles are generated snapshot by snapshot in AirSim. 
Furthermore, the trajectory of each vehicle is precisely aligned with AirSim in Wireless InSite.
For clarity, trajectories of vehicles under high VTDs in the urban crossroad are shown in Table~\ref{trajectories}.
\begin{table*}[]
\centering
\caption{Detailed Parameter Setting of Trajectories of Vehicles Under High VTDs}
\renewcommand\arraystretch{1.1}
\label{trajectories}
\begin{tabular}{cccccc}
\hline
\begin{tabular}[c]{@{}c@{}}Vehicle\\ name\end{tabular} & \begin{tabular}[c]{@{}c@{}}X-axis velocity\\ (m/snapshot)\end{tabular} & \begin{tabular}[c]{@{}c@{}}Y-axis velocity\\ (m/snapshot)\end{tabular} & \begin{tabular}[c]{@{}c@{}}Z-axis velocity\\ (m/snapshot)\end{tabular} & \begin{tabular}[c]{@{}c@{}}Movement\\ start snapshot\end{tabular} & \begin{tabular}[c]{@{}c@{}}Movement\\ end snapshot\end{tabular} \\ \hline
Car1 &0.0833&0&0&1&  300           \\ \hline
Car2  &-0.1 &0&0&1&  830            \\ \hline
Car3  &-0.1&0&0&1&     950        \\ \hline
Car4 &0&0.07&0&1&   1500          \\ \hline
Car5  &0&-0.15&0&1&  835           \\ \hline
Car6  &0&0.06&0&1& 1500        \\ \hline
S1  &0&0.04&0&1& 1500        \\ \hline
S2  &0 &0.03&0&1&  1500            \\ \hline
S3  &0&-0.05&0&1& 1500        \\ \hline
S4  &0&0.05&0&1&     1500       \\ \hline
S5 &0.05&0&0&1&  320          \\ \hline
S6  &0&-0.1&0&1&  480          \\ \hline
S7  &0&0.03&0&1&  1500         \\ \hline
S8  &0&-0.03&0&1& 1500        \\ \hline
S9  &-0.01&0&0&1& 1500        \\ \hline
S10  &0.01&0&0&1& 1500        \\ \hline
S11  &0.06&0&0&1&  1500      \\ \hline
S12  &0&0.02&0&1& 1500        \\ \hline
S13  &0&-0.03&0&1&  1500          \\ \hline
S14  &-0.02&0&0&1& 1500        \\ \hline
\end{tabular}
\end{table*}
Different cars are set with different trajectories, speeds, and durations of motion to construct complex and dynamic V2V scenarios.
Meanwhile, each vehicle is equipped with a LiDAR device and a communication unit. 
Positions of sensors and transceivers are precisely aligned with trajectories of the corresponding vehicles.
The LiDAR device in AirSim has $16$~channels and $10$~Hz scanning frequency. 
The communication unit in Wireless InSite is operated under a typical mmWave frequency band, i.e., $f_c = 28$~GHz carrier frequency with $2$~GHz communication bandwidth. 
Each communication unit is equipped with a transmitter (Tx) and a Rx. The numbers of antennas at Tx and Rx are $M_{\rm T} = M_{\rm R} = 1$. 
By adopting the RT technology, the electromagnetic propagation scatterer and channel impulse response (CIR) for 18 pairs of V2V links among form Car1 to Car6 under each VTD condition are obtained.

\section{Exploration of Mapping Relationship Between Physical Environment and Electromagnetic Space}
In this section, the preprocessing and environmental feature extraction of LiDAR point clouds are introduced, and a novel ScaR algorithm is further developed, which builds a bridge between LiDAR point clouds and scatterers.
On this basis, the mapping relationship between physical environment and electromagnetic space is explored for the first time.

\subsection{Environmental Feature Extraction}

The large amount of data in LiDAR point clouds results in significant computational complexity. 
Moreover, the noise and redundant points in raw LiDAR point clouds can significantly interfere with feature extraction and mapping relationship exploration.
Therefore, the preprocessing of LiDAR point clouds is introduced to eliminate redundant information and retain essential data within LiDAR point clouds.
Meanwhile, to characterize the electromagnetic propagation mechanism, the environmental feature related to scatterer spatial distribution is extracted. 

\subsubsection{Preprocessing of LiDAR Point Clouds}
\begin{figure*}[t]
\centering
\includegraphics[width=6.5in]{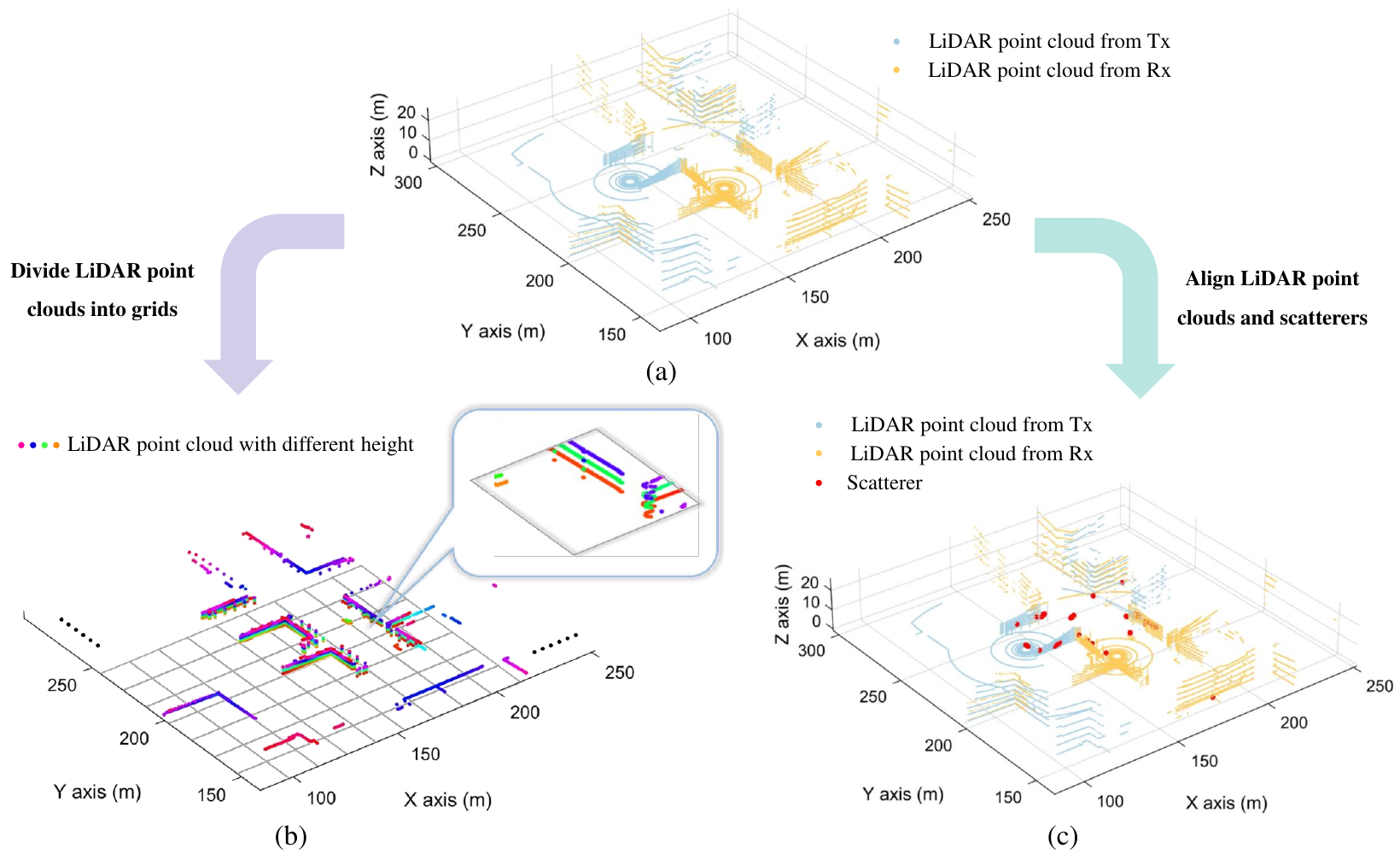}
\caption{Preprocessing and coordinate transformation of LiDAR point clouds.
(a) Combined LiDAR point clouds acquired from the Tx and Rx.
(b) LiDAR point cloud grid map.
(c) Alignment between LiDAR point clouds and scatterers.}
\label{LiDARfusion}
\end{figure*}

First, LiDAR point clouds acquired from the Tx and Rx are concatenated to facilitate the sharing of the observed physical environment, as shown in Fig.~\ref{LiDARfusion}(a).
The second step is to remove ground points generated from the laser reflection off the ground by setting a threshold $H_{\rm g}$. 
It is necessary because ground points typically constitute a significant portion of the LiDAR point clouds.
Meanwhile, ground points appear as a texture, which significantly interferes with LiDAR point cloud feature extraction.
After removing ground points, LiDAR point clouds contain effective environmental information associated with vehicles, buildings, trees, etc.
Then, we propose a novel height threshold filtering method, which further reduces redundancy in LiDAR point clouds.
Specifically, LiDAR point clouds with heights exceeding a predefined threshold $H_{\rm h}$, typically generated by reflections from tall buildings, are removed. 
This is because that electromagnetic propagation scatterers around ground vehicles are not generally distributed in the upper regions of buildings due to the path loss.
Finally, the VR method in~\cite{huang2024lidar}, which is widely used in channel modeling, is introduced to further reduce redundancy in LiDAR point clouds. 
Scatterers are visible and contribute to channel characterization only if they are within the VR. 
Similar to~\cite{huang2024lidar}, the VR is set to an ellipsoid with the Tx and the Rx as two focuses. 
The major axis $2a(t)$ of the ellipsoid can be calculated as the sum of the distances from the scatterers to the transceivers. 
The minor axis $2b(t)$ and the focal length $2c(t)$ of the ellipsoid is equal to the distance $D_{\rm tcv}(t)$ between transceivers.
Based on the VR method, the preprocessed LiDAR point clouds within the VR are filtered out and named \textit{valid LiDAR point clouds}, which are expressed as 
\begin{equation}
    \mathbb S(t)=\{\mathbf L(t)|\mathbf L(t)=[x_{\rm LiDAR}^{\rm v}(t),y_{\rm LiDAR}^{\rm v}(t),z_{\rm LiDAR}^{\rm v}(t)]\}.
\end{equation}

\subsubsection{Feature Extraction of LiDAR Point Clouds}
\begin{figure}[t]
\centering
\includegraphics[width=3.5in]{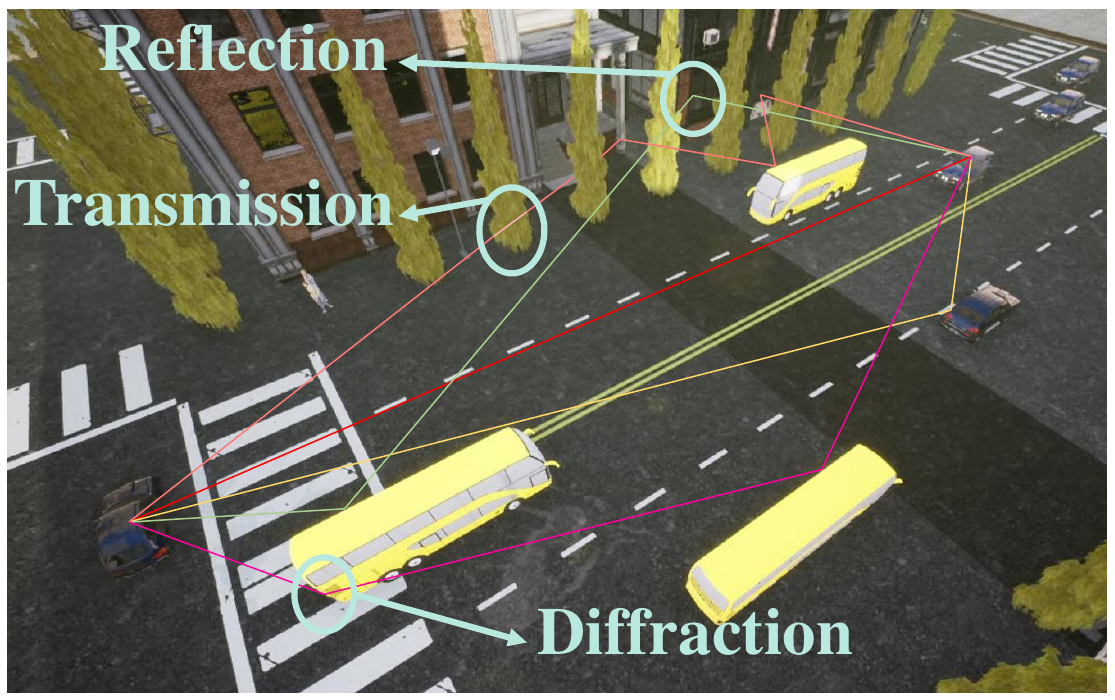}
\caption{RT results in electromagnetic space.}
\label{shexiantu}
\end{figure}

Scatterers between the transceivers can be accurately obtained by the deterministic channel modeling method, such as RT technology, in a constructed physical environment.
Therefore, the underlying mechanism of RT is embedded in the environmental feature extraction process of \textit{valid LiDAR point clouds}. 
The basic mechanism of RT technology involves emitting a series of rays from the Tx. 
Rays undergo processes, such as reflection, diffraction, and transmission, ultimately reaching the Rx, which are shown in Fig. \ref{shexiantu}.
During this process, the electromagnetic scatterers are determined.
Each ray is defined by its specific starting point, scatterers, direction, and energy. 
These attributes are crucial for accurately mimicking the propagation path and attenuation characteristics of electromagnetic waves.
To accurately mimic and obtain these ray properties, accurate 3D modeling of the environment is crucial.
This modeling incorporates geometric representations of various elements, such as buildings, trees, and terrain, and material properties including reflectivity and refractive index.
As a result, the primary factors influencing the distribution of scatterers include positions, materials, and angles of objects within the environment.

Based on the aforementioned analysis, features of \textit{valid LiDAR point clouds} are extracted in terms of density, height, and distance.
For this purpose, the widely used technique of grid feature extraction~\cite{Jiang_2020_CVPR} from computer vision and image processing can be leveraged.
This method involves dividing images or spatial data into regular grid units and extracting features from each unit.
The extreme values of \textit{valid LiDAR point clouds} in the $x$-axis and $y$-axis directions are given as $x_{\rm min}(t)$, $x_{\rm max}(t)$, $y_{\rm min}(t)$, $y_{\rm max}(t)$. 
The rectangular area determined by four coordinates, i.e., $(x_{\rm min}(t),y_{\rm min}(t))$, $(x_{\rm min}(t),y_{\rm max}(t))$, $(x_{\rm max}(t),y_{\rm min}(t))$, $(x_{\rm max}(t),y_{\rm max}(t))$, is evenly divided along the $x$-axis and $y$-axis.
A grid map with a dimension of $g_{\rm x} \times g_{\rm y}$ is formed and the length and width of each grid can be expressed as
\begin{equation}
    x_{\rm grid}(t)=\frac{(x_{\rm max}(t)-x_{\rm min}(t))}{g_{\rm x}}
\end{equation}
\begin{equation}
    y_{\rm grid}(t)=\frac{(y_{\rm max}(t)-y_{\rm min}(t))}{g_{\rm y}}.
\end{equation}
Then, each \textit{valid LiDAR point cloud} is dropped onto the corresponding grid based on the coordinate.
The Bird's eye view of the LiDAR point cloud collected by the transceiver in the $t$-th snapshot of the urban crossroad with high VTD is shown in Fig.~\ref{LiDARfusion}(b).
The black grid represents the LiDAR point cloud grid and LiDAR point clouds within one of the grids are displayed in the red box in detail. 
This grid is determined by four coordinates, i.e., $(x_{\rm min}^{\rm grid}(t),y_{\rm min}^{\rm grid}(t))$, $(x_{\rm min}^{\rm grid}(t),y_{\rm max}^{\rm grid}(t))$, $(x_{\rm max}^{\rm grid}(t),y_{\rm min}^{\rm grid}(t))$, $(x_{\rm max}^{\rm grid}(t),y_{\rm max}^{\rm grid}(t))$.
The LiDAR point clouds within this grid are given as 
\begin{equation}
\begin{aligned}
    \mathbb S^{\rm grid}(t)=\{\mathbf L(t)|x_{\rm min}^{\rm grid}(t) \leq x_{\rm LiDAR}^{\rm v}(t)<x_{\rm max}^{\rm grid}(t),\\
    y_{\rm min}^{\rm grid}(t) \leq y_{\rm LiDAR}^{\rm v}(t)<y_{\rm max}^{\rm grid}(t)\}.
\end{aligned}
\end{equation}

Furthermore, take this grid as an example for feature extraction on LiDAR point clouds.
First, the density of LiDAR point clouds in this grid is calculated as
\begin{equation}
    \rho_{\rm grid}(t)=\frac{|\mathbb S^{\rm grid}(t)|}{x_{\rm grid}(t)\times y_{\rm grid}(t)}
\end{equation}
where $|\mathbb S^{\rm grid}(t)|$ denotes the number of elements in $\mathbb S^{\rm grid}(t)$. 
Next, the maximum height of the LiDAR point clouds in this grid is extracted, which can be calculated as
\begin{equation}
\begin{aligned}
    &h_{\rm grid}(t)=\max_{L(t)\in \mathbb S^{\rm grid}(t)} z_{\rm LiDAR}^{\rm v}(t).
    % \\
    % &\mathbf L(t)=[x_{\rm LiDAR}^{\rm v}(t),y_{\rm LiDAR}^{\rm v}(t),z_{\rm LiDAR}^{\rm v}(t)]\in \mathbb S^{\rm grid}(t).
\end{aligned}
\end{equation}
The maximum height represents the 3D information of the physical environment. 
Finally, a position encoding layer, which contains the distance from the transceiver to LiDAR point clouds in this grid, is introduced as scatterers become sparser as the distance increases.
For grids containing LiDAR point clouds, the distance is calculated as
\begin{equation}
    d_{\rm grid}(t)=\frac{d_{\rm Tg}+d_{\rm Rg}}{2}
\end{equation}
where $d_{\rm Tg}$ denotes the distance between the center of the grid and Tx and $d_{\rm Rg}$ denotes the distance between the center of the grid and Rx.
To avoid interference with ScaR, values in the position encoding layer for grids without LiDAR point clouds are uniformly set to the maximum value in $d_{\rm grid}(t)$.

\subsection{Scatterer Recognition Based on LiDAR Point Cloud Grid Map}
With the help of the LiDAR point clouds, environmental features of the physical environment are obtained.
Furthermore, based on the obtained environmental feature, a novel ScaR algorithm is developed for the first time.

\subsubsection{Coordinate Transformation of LiDAR Point Clouds} The coordinate system, where the original LiDAR point clouds are located, follows a left-hand system with the $x$-axis aligned with the forward direction of the vehicle and the $z$-axis pointing upwards vertically. 
However, scatterers are positioned in the world coordinate system.
The coordinate of the LiDAR is given as $\mathbf R(t)=[x_{\rm L}(t),y_{\rm L}(t),z_{\rm L}(t)]$. 
The relative coordinate of the \textit{valid LiDAR point cloud} is given as $\mathbf L(t)=[x_{\rm LiDAR}^{\rm v}(t),y_{\rm LiDAR}^{\rm v}(t),z_{\rm LiDAR}^{\rm v}(t)]$.
When the angle between the forward direction of the vehicle and the $x$-axis of the world coordinate system is $\theta$, the absolute coordinate of the \textit{valid LiDAR point cloud} is calculated as
\begin{equation}
\begin{aligned}
    L^{\rm a}(t)=[&x_{\rm LiDAR}^{\rm v}(t)\cos \theta+y_{\rm LiDAR}^{\rm v}(t)\sin \theta+x_{\rm L}(t),\\
    &x_{\rm LiDAR}^{\rm v}(t)\sin \theta-y_{\rm LiDAR}^{\rm v}(t)\cos \theta-y_{\rm L}(t),\\
    &-z-z_{\rm L}(t)
    ].
\end{aligned}
\end{equation}
In Fig.~\ref{LiDARfusion}(c), LiDAR point clouds and scatterers between Car1 and Car3 under low VTD are depicted, where Car1 is the Tx and Car3 is the Rx.
Scatterers are represented by red circles. It can be observed that scatterers are located on LiDAR point clouds thus validating the precise alignment between physical environment and electromagnetic space.
Note that the aforementioned processing steps, including preprocessing, feature extraction, and coordinate matching, are all based on the simulation results in Wireless InSite. 
The following will employ the processed data to directly recognize scatterers within LiDAR point clouds through data-driven methods.

\subsubsection{Scatterer Recognition Network Design}
Since there is a significant difference in data representation and acquisition frequency between LiDAR point clouds and scatterers, it is difficult to explore the mapping relationship by mathematical expressions.
Therefore, it is necessary to adequately leverage advantages of ANNs to explore complex nonlinear mapping relationships.

\begin{figure*}[t]
\centering
\includegraphics[width=7in]{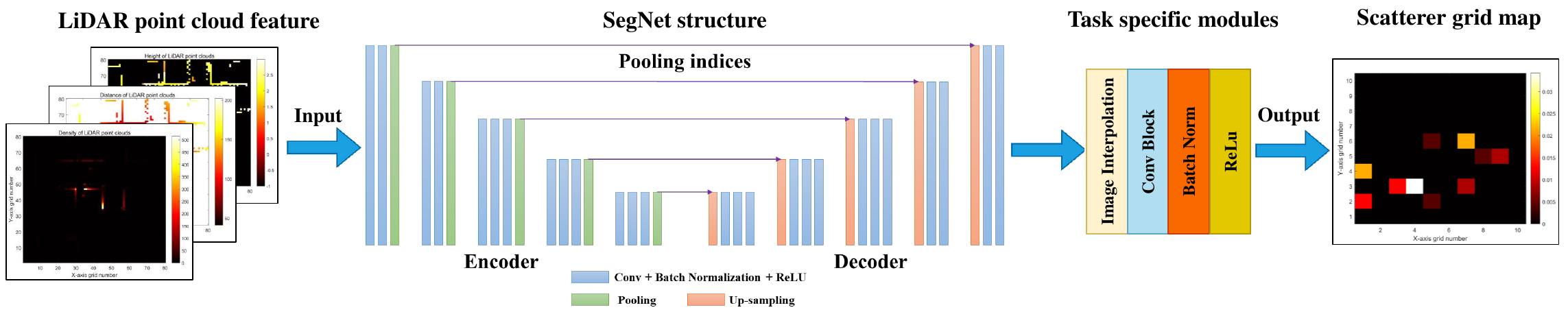}
\caption{Customized network that builds on the SegNet architecture.}
\label{network-structure}
\end{figure*}

To recognize scatterers from LiDAR point clouds, we employ a customized network, which builds on the SegNet architecture, as shown in Fig. \ref{network-structure}. 
The input, output, and structure of the customized network are given below.

First, the three-layer image related to environmental features extracted from LiDAR point clouds is utilized as the input of the network.
The first layer is the density grid map of LiDAR point clouds, which provides a visual depiction of the scenario layout. The distribution of buildings, vehicles, trees, etc., closely associated with the distribution of scatterers is indicated.
The second layer is the height encoding grid map of LiDAR point clouds, which supplements the 3D information of the physical environment on the basis of the first layer.
The third layer is the position encoding layer, where the distance between LiDAR point clouds and transceivers is indicated. 
Note that the position encoding layer further enhances the network's understanding of the impact of distance on the distribution of scatterers. 

Second, the network architecture is introduced, including the backbone and task specific module. 
SegNet is selected as the backbone, which is a fully convolutional encoder-decoder, designed to perform a pixel-by-pixel transformation of an input image to a target output image through supervised learning. 
On the one hand, fully convolutional networks (FCNs) utilize shared weights across the entire input, which significantly reduces the number of parameters compared to fully connected networks. 
This parameter efficiency results in greater computational efficiency of FCNs and enables training on larger datasets.
On the other hand, FCNs can incorporate multi-scale information through the use of pooling and upsampling layers. 
This enables them to capture both local details and global context in the input, leading to enhanced performance in tasks, such as object detection and image segmentation.

The SegNet architecture includes an encoder and a decoder.
The encoder is composed of 13 convolutional layers and 5 pooling layers and the decoder is composed of 13 convolutional layers and 5 up-sampling layers.
The encoder aims to extract semantic information and high-level features from the image via hierarchical abstraction. The output of the encoder is a series of feature maps with varying levels of abstraction.
The decoder's objective is to restore details and spatial structures of the image by combining feature maps from the encoder with high-resolution detail information.

Let $O_{\rm e}^i,i=1,2,\cdots ,5,$ be the output of the $i$-th pooling layer, which can be calculated as 
\begin{equation}
    O_{\rm e}^i=\zeta_{\rm pool}^{{\rm e},i}(\varphi_{\rm conv}^{{\rm e},i}(I_{\rm e}^i))
\end{equation}
where $I_{\rm e}^i,i=1,2,\cdots ,5,$ is the input of the first convolutional layer in each part of the encoder, respectively.
$\zeta_{\rm pool}^{{\rm e},i}(\cdot),i=1,2,\cdots ,5,$ and $\varphi_{\rm conv}^{{\rm e},i}(\cdot),i=1,2,\cdots ,5,$ represent the operation of the pooling layer and multiple convolutional layers, respectively. 
It can be observed that SegNet passes the pooling indices from the compression path in the encoder to the expression path in the decoder. 
This structure is named skip connections, enabling less memory to be occupied while keeping the topological information in the decoder.
For our ScaR task, the topological information of the physical environment extracted from LiDAR point clouds is significantly important to be captured.
The output of the last convolutional layer in each part of the decoder can be written as 
\begin{equation}
    O_{\rm d}^i=\varphi_{\rm conv}^{{\rm d},i}(\psi_{\rm up}^{{\rm d},i} (I_{\rm d}^i)), \zeta_{\rm pool}^{{\rm e},5-i}(\cdot)\mapsto \psi_{\rm up}^{{\rm d},i}(\cdot)
\end{equation}
where $I_{\rm d}^i,i=1,2,\cdots ,5,$ is the input of the up-sampling layer in each part of the decoder, respectively.
$\psi_{\rm up}^{{\rm d},i}(\cdot),i=1,2,\cdots ,5,$ and $\varphi_{\rm conv}^{{\rm e},i}(\cdot),i=1,2,\cdots ,5,$ represent the operation of the up-sampling layer and multiple convolutional layers, respectively.
`$\mapsto$' represents the transfer of pooling indices from the encoder to the decoder.

In addition, SegNet was originally designed for image semantic segmentation, which possesses a significant difference from our task.
In image semantic segmentation, the predicted values are discrete and represent certain classification classes.
However, in the ScaR task, the values are continuous and represent the continuous scatterer density values.
In the original structure of SegNet, the last layer generally consists of a $k$-class softmax classifier.
Moreover, the original output of SegNet is an image with the same dimension as the input.
However, the dimensions of our task's output, i.e., $s_{\rm x}\times s_{\rm y}$, are defined based on actual needs.
To address aforementioned problems, the $k$-class softmax classifier is removed in the customized network, and a task specific module is adopted after the SegNet structure. 
This modification allows us to tailor the network's output and incorporate specific functionalities relevant to the ScaR task.
Specifically, the task specific module includes an image interpolation layer, convolutional layer, batch normalization layer, and activation function.
To convert the dimension of each channel of the output tensor to $s_{\rm x}\times s_{\rm y}$, the image interpolation layer is implemented by the F.interpolate function in PyTorch, which is expressed as
\begin{equation}
    O_{\rm t}^{s_{\rm x}\times s_{\rm y}}=\Gamma_{\rm ip}(O_{\rm d}^{g_{\rm x}\times g_{\rm y}})
\end{equation}
where $O_{\rm d}^{g_{\rm x}\times g_{\rm y}}$ is the output of the decoder of SegNet, and $O_{\rm t}^{s_{\rm x}\times s_{\rm y}}$ is the output of the image interpolation layer.
Then, the convolutional layer is exploited to convert tensor $O_{\rm d}^{s_{\rm x}\times s_{\rm y}}$ from multi-channel to single channel.
Finally, the output of the customized network is expressed as 
\begin{equation}
\label{relu}
    O_{\rm f}=f_{\rm ReLU}(\varrho_{\rm bn}(\varphi_{\rm conv}(O_{\rm t}^{s_{\rm x}\times s_{\rm y}})))
\end{equation}
where $\varrho_{\rm bn}(\cdot)$ represents the operation of the batch normalization layer, and $f_{\rm ReLU}(\cdot)$ represents the ReLU activation function, which can be expressed as $f_{\rm ReLU}(x)=\max (0,x)$. 
ReLU addresses the vanishing gradient problem, which can occur in deep neural networks when gradients become extremely small during backpropagation.

Third, the output of the customized network is a single-layer image related to the spatial distribution of scatterers.
To be specific, the rectangular area determined by four coordinates, i.e., $(x_{\rm min}(t),y_{\rm min}(t))$, $(x_{\rm min}(t),y_{\rm max}(t))$, $(x_{\rm max}(t),y_{\rm min}(t))$, $(x_{\rm max}(t),y_{\rm max}(t))$, is evenly divided along the $x$-axis and $y$-axis to form a grid map with a dimension of $s_{\rm x} \times s_{\rm y}$.
The length and width of each grid can be expressed as
\begin{equation}
    x_{\rm grid}^{\prime}(t)=\frac{(x_{\rm max}(t)-x_{\rm min}(t))}{s_{\rm x}}
\end{equation}
\begin{equation}
    y_{\rm grid}^{\prime}(t)=\frac{(y_{\rm max}(t)-y_{\rm min}(t))}{s_{\rm y}}.
\end{equation}
Then, scatterers within the VR are dropped onto the corresponding grid based on coordinates. 
It is noteworthy that scatterers, similar to LiDAR point clouds, are filtered by VR.
Let $\mathbb R^{\rm grid}(t)$ be a point set corresponding to one of the scatterer girds.
The density of scatterers in this grid is calculated as
\begin{equation}
    \xi_{\rm grid}(t)=\frac{|\mathbb R^{\rm grid}(t)|}{x_{\rm grid}^{\prime}(t)\times y_{\rm grid}^{\prime}(t)}.
\end{equation}
Note that the output of ReLU in (\ref{relu}) is non-negative, which is the same as the scatterer density.

\section{Multi-Modal Intelligent V2V Channel Modeling Based on Scatterer Recognition}

With the help of LiDAR point clouds and the mapping relationship between physical environment and electromagnetic space, the spatial distribution of scatterers is recognized. 
Based on the recognized scatterers, a novel multi-modal intelligent V2V channel model for sensing-communication integration is proposed.

\subsection{Processing of Recognized Scatterer Grid Maps}
\begin{figure}[t]
\centering
\includegraphics[width=3.5in]{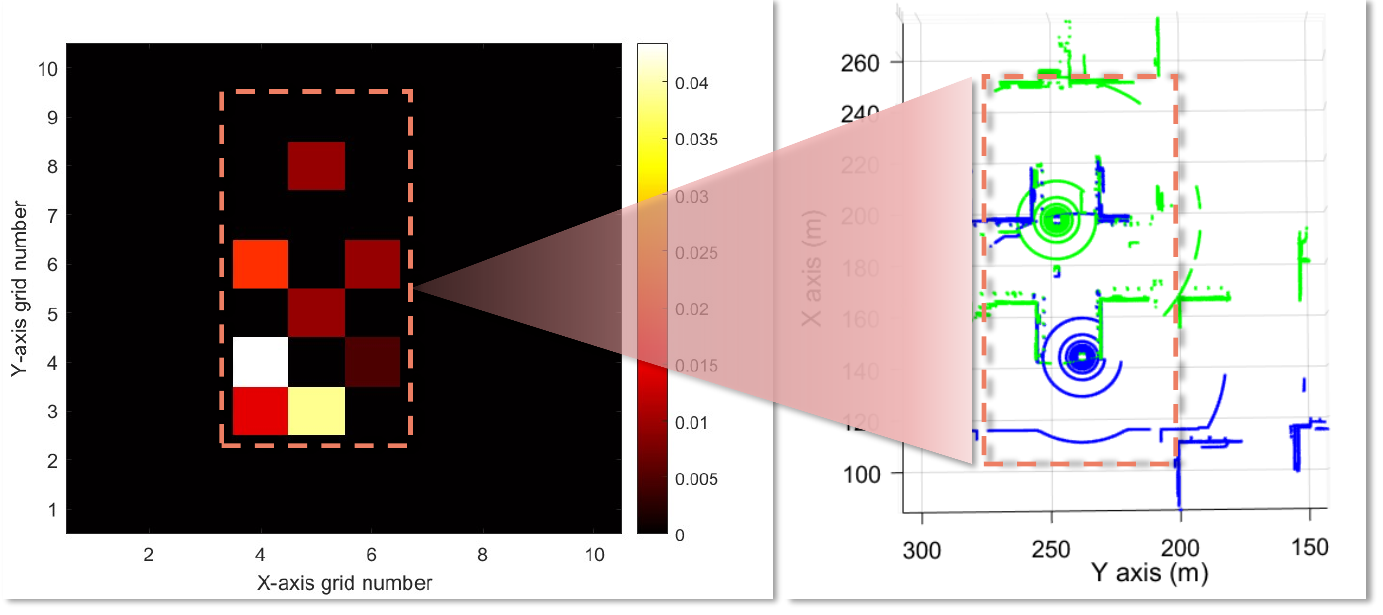}
\caption{Matching of predicted scatterer grid map and VR.}
\label{scattererfeedback}
\end{figure}

\begin{figure}[t]
\centering
\includegraphics[width=3.2in]{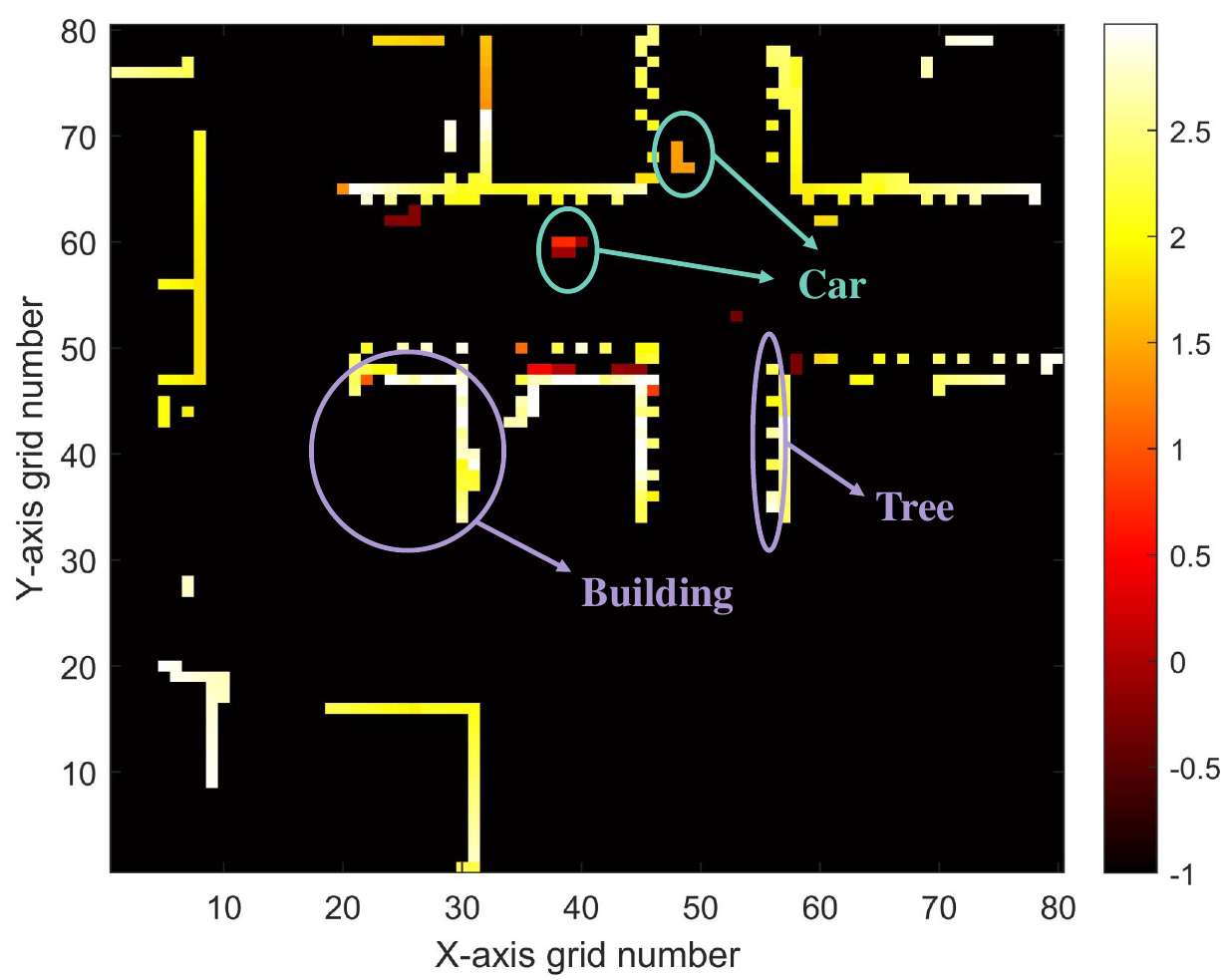}
\caption{LiDAR point cloud height grid map under medium VTDs at time $t$.}
\label{LiDARheight}
\end{figure}

In existing channel models, the modeling of precise scatterer positions and a large number of random parameters bring huge challenges.
Instead of randomly generating scatterers, we exploit the predicted scatterer grid map $O_{\rm f}$ to generate the position of scatterers.
The predicted scatterer density $\hat{\xi}_{\rm grid}(t)$ of each grid is first converted to the scatterer number, which is calculated as 
\begin{equation}
\hat{\kappa}_{\rm grid}(t)={\rm round}(\hat{\xi}_{\rm grid}(t)x_{\rm grid}^{\prime}(t) y_{\rm grid}^{\prime}(t))
\end{equation}
where ${\rm round}(\cdot)$ represents the rounding operation.
To align the scatterer density information with the corresponding physical environment, the predicted scatterer grid map and VR are then matched, as illustrated in Fig. \ref{scattererfeedback}.
According to the number of scatterers within each grid, we randomly generate $\hat{\kappa}_{\rm grid}(t)$ scatterers in each grid.
Subsequently, scatterers are divided into three categories, i.e., static scatterers, dynamic scatterers, and unknown scatterers.
If all the values in point cloud density grids within a scatterer grid are zero, scatterers in that grid are classified as unknown scatterers.
Furthermore, to distinguish between dynamic and static scatterers, LiDAR point cloud height grids are considered.
The corresponding value for each remaining scatterer in the point cloud height grid is denoted as $h_{\rm grid}^{\rm s}(t)$.
Dynamic scatterers are identified as scatterers with $h_{\rm grid}^{\rm s}(t)$ below a threshold $h_{\rm th}$, while static scatterers are identified as scatterers with $h_{\rm grid}^{\rm s}(t)$ above a threshold $h_{\rm th}$.
For example, Fig. \ref{LiDARheight} shows a LiDAR point cloud height grid map under medium VTDs at time $t$.
It can be observed that there is a significant height difference between static objects, such as buildings and trees, and dynamic vehicles.
Consequently, a reasonable threshold $h_{\rm th}$ of 1.5 m can be set.

According to scatterer coordinates generated by the predicted scatterer grid map, the number, distance, and angle information can be obtained. 
In addition, the path power is an exponential function of the path delay~\cite{ii2007winner}.
The power of the $k$-th static path and $l$-th dynamic path in the communication link is denoted by
\begin{equation}
    P_{k}^{\rm s}(x)=\exp (-\chi _{\rm s} \tau _{k}(t)-\nu _{\rm s})10^{-\frac{\rho _{\rm s}}{10}}
\end{equation}
\begin{equation}
    P_{l}^{\rm d}(x)=\exp (-\chi _{\rm d} \tau _{l}(t)-\nu _{\rm d})10^{-\frac{\rho _{\rm d}}{10}}
\end{equation}
where $\chi _{\rm s/d}$ and $\nu _{\rm s/d}$ are parameters of static/dynamic scatterers related to delay.
$\tau _{k/l}$ is the delay of the $k$/$l$-th static/dynamic path.
Furthermore, $\rho _{\rm s}$ and $\rho _{\rm d}$ obey Gaussian
distributions.

\subsection{Framework of the Proposed Channel Model}
\begin{figure}[t]
\centering
\includegraphics[width=3.5in]{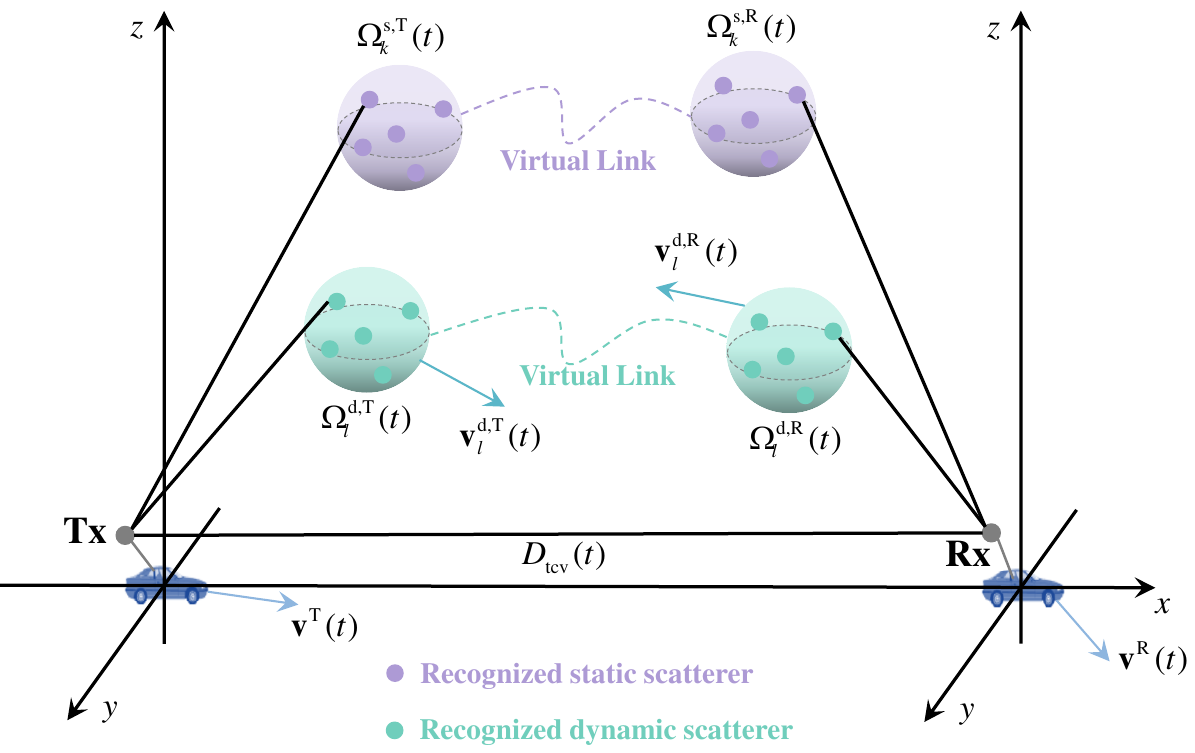}
\caption{Geometry of the proposed multi-modal intelligent V2V channel model.}
\label{channelmodel}
\end{figure}

The geometric representation of the proposed channel model is depicted in Fig. \ref{channelmodel}.
The transceiver distance is $D_{\rm tcv}(t_0)$ at the initial time. The center frequency and bandwidth of communications are 28 GHz and 2 GHz, respectively.
Numbers of static and dynamic scatterers, i.e., $I_{\rm s}(t_0)$ and $I_{\rm d}(t_0)$, at the initial time are obtained through the predicted scatterer grid map.
Then, recognized static and dynamic scatterers are clustered via density-based spatial clustering of applications with noise (DBSCAN) to form static and dynamic clusters.
The cluster whose centroid is closer to Tx/Rx is defined as the Tx/Rx cluster.
Furthermore, the Tx cluster and the Rx cluster are randomly shuffled and matched to obtain the twin-cluster.
The propagation between twin clusters is abstracted as a virtual link, where other clusters may exist and introduce the second-order and beyond reflection and scattering components.
The $k$-th Tx/Rx static cluster and the $j$-th Tx/Rx dynamic cluster are denoted as $\Omega _k^{\rm s,T/R}$ and $\Omega _l^{\rm d,T/R}$. 
$\mathbf{v}_l^{\rm d,T/R}(t)$ represents the velocity vector of the dynamic cluster. 
Finally, velocity vectors of Tx and Rx are $\mathbf{v}^{\rm T}(t)$ and $\mathbf{v}^{\rm R}(t)$, respectively.

\subsubsection{LoS Component}
The complex channel gain of the LoS component can be represented as 
\begin{equation}
    h^{\rm LoS}(t)=R(t)\exp[j2\pi \int _{t_0}^t f_{\rm d}^{\rm LoS}(t){\rm d}t+j\varrho^{\rm LoS}(t)]
\end{equation}
where $R(t)$ represents a rectangular window function~\cite{gutierrez2017non}, which can be denoted by
\begin{equation}
    R(t)=\begin{cases}
1,& t_0\leq t \leq T_0,\\ 
0,& otherwise.
\end{cases}
\end{equation}
The Doppler frequency $f_{\rm d}^{\rm LoS}(t)$, phase shift $\varrho^{\rm LoS}(t)$, as well as delay $\tau ^{\rm LoS}(t)$ can be given by
\begin{equation}
    f_{\rm d}^{\rm LoS}(t)=\frac{\langle D_{\rm tcv}(t), \mathbf{v}^{\rm T}-\mathbf{v}^{\rm R}\rangle}{\lambda \|D_{\rm tcv}(t)\|}
\end{equation}
\begin{equation}
    \varrho^{\rm LoS}(t)=\varrho_0+\frac{2\pi}{\lambda}\|D_{\rm tcv}(t)\|
\end{equation}
\begin{equation}
    \tau ^{\rm LoS}(t)=\frac{\|D_{\rm tcv}(t)\|}{c}
\end{equation}
where $\langle \cdot\rangle$, $\varrho_0$, $\lambda$, and $c$ represent the inner product, initial phase shift, carrier wavelength, and speed of light, respectively.

\subsubsection{Ground Reflection Component}
The complex channel gain of the ground reflection component can be represented as 
\begin{equation}
\begin{aligned}
    h^{\rm GR}(t)=&R(t)\sqrt{P^{\rm GR}(t)}
    \exp\{j2\pi[\int _{t_0}^t  f^{\rm GR,T} (t) {\rm d}t \\&+\int _{t_0}^t  f^{\rm GR,R} (t) {\rm d}t 
    ]+j\varrho^{\rm GR}(t)
    \}
\end{aligned}
\end{equation}
where $P^{\rm GR}(t)$, $f^{\rm GR,T/R} (t)$, and $\varrho^{\rm GR}(t)$ denote the power, Doppler frequency at Tx/Rx, and phase shift of ground reflection component, respectively.

\subsubsection{NLoS Component via Dynamic Clusters}
The complex channel gain via the static twin-cluster $\Omega _k^{\rm s,T/R}$ by the $l_j$-th static scatterer can be denoted by
\begin{equation}
\begin{aligned}
    h_{l,l_j}^{\rm d}(t)=&R(t)\sqrt{P_{l,l_j}^{\rm d}(t)}
    \exp\{j2\pi[\int _{t_0}^t  f_{l,l_j}^{\rm d,T} (t) {\rm d}t \\&+\int _{t_0}^t  f_{l,l_j}^{\rm d,R} (t) {\rm d}t 
    ]+j\varrho_{l,l_j}^{\rm d}(t)
    \}
\end{aligned}
\end{equation}
where $P_{l,l_j}^{\rm d}(t)$ represents the normalized dynamic scatterer power. 
The Doppler frequency is computed by
\begin{equation}
    f_{l,l_j}^{\rm d,T/R} (t)=\frac{\langle D_{l,l_j}^{\rm d,T/R} (t), \mathbf{v}^{\rm T/R}(t)-\mathbf{v}_l^{\rm d,T/R}(t)\rangle}{\lambda \|D_{l,l_j}^{\rm d,T/R} (t)\|}
\end{equation}
where $D_{l,l_j}^{\rm d,T/R} (t)$ represents the distance between the Tx/Rx and the $l_j$-th static scatterer via the static cluster $\Omega _l^{\rm d,T/R}$.
The delay can be written by
\begin{equation}
    \tau _{l,l_j}^{\rm d}(t)=\frac{\|D_{l,l_j}^{\rm d,T} (t)\|+\|D_{l,l_j}^{\rm d,R} (t)\|}{c}+\tau _j^{\prime}(t)
\end{equation}
where $\tau _j^{\prime}(t)$ denotes the abstracted delay of the
virtual link between the static twin-cluster $\Omega _l^{\rm d,T}$ and $\Omega _l^{\rm d,R}$, which follows the exponential distribution.
The phase shift is computed by
\begin{equation}
    \varrho_{l,l_j}^{\rm d}(t)=\varrho_0+\frac{2\pi}{\lambda}[\|D_{l,l_j}^{\rm d,T} (t)\|+\|D_{l,l_j}^{\rm d,R} (t)\|+c\tau _j^{\prime}(t)].
\end{equation}

\subsubsection{NLoS Component via Static Clusters}
The complex channel gain via the static twin-cluster $\Omega _k^{\rm s,T/R}$ by the $k_i$-th static scatterer can be denoted by
\begin{equation}
\begin{aligned}
    h_{k,k_i}^{\rm s}(t)=&R(t)\sqrt{P_{k,k_i}^{\rm s}(t)}
    \exp\{j2\pi[\int _{t_0}^t  f_{k,k_i}^{\rm s,T} (t) {\rm d}t \\&+\int _{t_0}^t  f_{k,k_i}^{\rm s,R} (t) {\rm d}t 
    ]+j\varrho_{k,k_i}^{\rm s}(t)
    \}
\end{aligned}
\end{equation}
where $P_{k,k_i}^{\rm s}(t)$ represents the normalized static scatterer power. 
The Doppler frequency is computed by
\begin{equation}
    f_{k,k_i}^{\rm s,T/R} (t)=\frac{\langle D_{k,k_i}^{\rm s,T/R} (t), \mathbf{v}^{\rm T/R}(t)\rangle}{\lambda \|D_{k,k_i}^{\rm s,T/R} (t)\|}
\end{equation}
where $D_{k,k_i}^{\rm s,T/R} (t)$ represents the distance between the Tx/Rx and the $k_i$-th static scatterer via the static cluster $\Omega _k^{\rm s,T/R}$.
The delay can be written by
\begin{equation}
    \tau _{k,k_i}^{\rm s}(t)=\frac{\|D_{k,k_i}^{\rm s,T} (t)\|+\|D_{k,k_i}^{\rm s,R} (t)\|}{c}+\tau _i^{\prime}(t)
\end{equation}
where $\tau _i^{\prime}(t)$ denotes the abstracted delay of the
virtual link between the static twin-cluster $\Omega _k^{\rm s,T}$ and $\Omega _k^{\rm s,R}$, which follows the exponential distribution.
The phase shift is computed by
\begin{equation}
    \varrho_{k,k_i}^{\rm s}(t)=\varrho_0+\frac{2\pi}{\lambda}[\|D_{k,k_i}^{\rm s,T} (t)\|+\|D_{k,k_i}^{\rm s,R} (t)\|+c\tau _i^{\prime}(t)].
\end{equation}

Overall, the CIR of the proposed multi-modal intelligent V2V 
channel model can be denoted by
\begin{equation}
\begin{aligned}
    h(t,\tau)=&\sqrt{\frac{\Upsilon(t)}{\Upsilon(t)+1}}h^{\rm LoS}(t)\delta (\tau-\tau^{\rm LoS}(t))
    \\+&\sqrt{\frac{\Theta^{\rm GR}(t)}{\Upsilon(t)+1}}h^{\rm GR}(t)\delta (\tau-\tau^{\rm GR}(t))
    \\+&\sqrt{\frac{\Theta^{\rm s}(t)}{\Upsilon(t)+1}}\sum _{k=1}^{J_{\rm s}(t)}\sum _{k_i=1}^{I_{\rm s}(t)}h_{k,k_i}^{\rm s}(t)\delta (\tau-\tau_{k,k_i}^{\rm s}(t))
    \\+&\sqrt{\frac{\Theta^{\rm d}(t)}{\Upsilon(t)+1}}\sum _{l=1}^{J_{\rm d}(t)}\sum _{l_j=1}^{I_{\rm d}(t)}h_{l,l_j}^{\rm d}(t)\delta (\tau-\tau_{l,l_j}^{\rm d}(t))
\end{aligned}
\end{equation}
where $\Upsilon(t)$ represents the Ricean factor.
$\Theta^{\rm GR}(t)$, $\Theta^{\rm s}(t)$, and $\Theta^{\rm d}(t)$ represent the power ratios of the ground reflection component, component via static clusters, and component resulting from dynamic clusters, which satisfy $\Theta^{\rm GR}(t)+\Theta^{\rm s}(t)+\Theta^{\rm d}(t)=1$.

\subsection{Capturing of Channel Non-Stationarity and Consistency}

Based on the proposed multi-modal intelligent V2V channel model, channel non-stationarity and consistency in the time and frequency domains are mimicked. 

As time evolves, the physical environment and transceiver positions undergo changes, and thus distinct LiDAR point clouds are captured at different time instants. 
Through the ScaR algorithm based on the tight mapping relationship between physical environment and electromagnetic space, scatterers evolve alongside time varying LiDAR point clouds. 
Consequently, channel non-stationarity in the time domain is captured.
Additionally, since the vehicular movement and environmental changes at adjacent instants are continuous, LiDAR point clouds acquired from the physical environment exhibit similarities. 
As a result, the recognized scatterers from these LiDAR point clouds also demonstrate similarities at adjacent instants, thus capturing channel consistency in the time domain.

The deployment of mmWave technology requires the capturing of frequency non-stationarity~\cite{huang2024lidar}.
A frequency-dependent factor $(\frac{f}{f_c})^\vartheta$ is incorporated into the time-varying transfer function (TVTF). 
The TVTF is derived by using the Fourier transform to CIR $h(t,\tau)$ in respect of delay $\tau$~\cite{bai2022non}, which is denoted by
\begin{equation}
    H(t,f)=\int _{-\infty}^{\infty}h(t,\tau)\exp (-j2\pi f\tau){\rm d}\tau.
\end{equation}
The TVTF with the frequency-dependent factor can be expressed as 
\begin{equation}
\begin{aligned}
    &H(t,f)=\sqrt{\frac{\Upsilon(t)}{\Upsilon(t)+1}}h^{\rm LoS}(t)\exp{[-j2\pi f\tau^{\rm LoS}(t)]}
    \\&+\sqrt{\frac{\Theta^{\rm GR}(t)}{\Upsilon(t)+1}}(\frac{f}{f_c})^\vartheta h^{\rm GR}(t)\exp{[-j2\pi f\tau^{\rm GR}(t)]}
    \\&+\sqrt{\frac{\Theta^{\rm s}(t)}{\Upsilon(t)+1}}(\frac{f}{f_c})^\vartheta\sum _{k=1}^{J_{\rm s}(t)}\sum _{k_i=1}^{I_{\rm s}(t)} h_{k,k_i}^{\rm s}(t)\exp{[-j2\pi f\tau_{k,k_i}^{\rm s}(t)]}
    \\&+\sqrt{\frac{\Theta^{\rm d}(t)}{\Upsilon(t)+1}}(\frac{f}{f_c})^\vartheta\sum _{l=1}^{J_{\rm d}(t)}\sum _{l_j=1}^{I_{\rm d}(t)} h_{l,l_j}^{\rm d}(t)\exp{[-j2\pi f\tau_{l,l_j}^{\rm d}(t)]}
\end{aligned}
\end{equation}
where $\vartheta$ is the frequency-dependent parameter that relies on the environment~\cite{molisch2005comprehensive}. 
By incorporating the frequency-dependent factor, the frequency-dependent path gain is replicated, thus modeling frequency non-stationarity.

\section{Channel Statistical Properties}
Key channel statistical properties, i.e.,  time-frequency correlation function (TF-CF) and Doppler power spectral density (DPSD), are derived.

\subsection{Time-Frequency Correlation Function}

Based on the TVTF, the TF-CF can be calculated by
\begin{equation}\label{TF-CF}
    \Lambda(t,f;\Delta t,\Delta f)=\mathbb E[H^{\rm *}(t,f)H(t+\Delta t,f+\Delta f)]
\end{equation}
where $\mathbb E[\cdot]$ and $(\cdot)^{\rm *}$ represent the expectation operation and complex conjugate operation, respectively.
Furthermore, the TF-CF of LoS components, ground reflection components, and NLoS components resulting from static and dynamic clusters can be calculated similarly to~\cite{huang2024lidar}.
Based on the TF-CF, the time auto-correlation function (TACF) and the frequency correlation function (FCF) can be derived by setting $\Delta f=0$ and $\Delta t=0$, respectively.

\subsection{Doppler Power Spectral Density}
The DPSD can be derived by taking the Fourier transform of the TACF, which is denoted by
\begin{equation}\label{DPSD}
    \Xi(t;f_{\rm D})=\int _{-\infty}^{+\infty} \Lambda(t;\Delta t)e^{-j2\pi f_{\rm D}\Delta t}{\rm d}(\Delta t)
\end{equation}
where $\Lambda(t;\Delta t)$ is the TACF and $f_{\rm D}$ is the Doppler frequency.

\section{Experimental Setup and Performance Evaluation}
This section introduces the evaluation dataset and metrics, as well as the network training methodology of the ScaR.
Then, simulation and verification of the proposed channel model are conducted by simulating and comparing key channel statistical properties with RT-based results.
Unless otherwise specified, simulations are conducted using the following parameter.
The carrier frequency is $f_{\rm c}=28$~GHz with 2~GHz communication bandwidth.
The numbers of antennas at Tx and Rx are $M_{\rm T}=M_{\rm R}=1$.

\subsection{Network Training and Validation}
\subsubsection{Dataset Overview}
The simulation V2V dataset under each VTD contains 27,000 samples, which consist of six pairs of transceivers for each of three street types in 1500 snapshots.
In each sample, the three-layer LiDAR point cloud feature grid maps serve as the network input, while the single-layer scatterer grid map is utilized as the network output.
The dataset under each VTD is divided into the training set, validation set, and test set in the proportion of 3 : 1 : 1.
After training the network on the training set, the performance of model is evaluated and hyper-parameters are optimized using the validation set. Subsequently, the performance testing is conducted on the network using the test set to evaluate the generalization ability of model ability on unprecedented data.

\subsubsection{Network Configuration and Hyper-parameters}
Table \ref{parameter1} shows hyper-parameters for the customized network design, including the size of LiDAR point cloud feature grid maps, the size of scatterer grid map, and architecture details of the ScaR network.
\begin{table}[t]
\centering
\renewcommand\arraystretch{1}
\caption{Hyper-Parameter for Network Design}
\label{parameter1}
\begin{tabular}{c|c}
\toprule 
\textbf{Parameter  }       & \textbf{Value} \\ \midrule
\begin{tabular}[c]{@{}c@{}}Size of LiDAR point cloud feature  \\ grid maps@$[3, g_{\rm x}, g_{\rm y}]$ \end{tabular}         &  [3, 80, 80]     \\ \midrule
Size of scatterer grid maps@$[1, s_{\rm x}, s_{\rm y}]$  &  [1, 10, 10]      \\ \midrule
\begin{tabular}[c]{@{}c@{}}Size of convolution kernels \\ in each part of encoder@$kernel\_size$\end{tabular} & (3, 3)      \\ \midrule
\begin{tabular}[c]{@{}c@{}}Size of convolution kernels \\ in each part of decoder@$kernel\_size$\end{tabular} &   (3, 3)     \\ \midrule
Parameters of image interpolation layer@$size$  &   (10, 10) \\ \midrule
\begin{tabular}[c]{@{}c@{}}Size of convolution kernel \\ in task specific module@$kernel\_size$\end{tabular}  &  (1, 1) \\ \bottomrule 
\end{tabular}
\end{table}
Table \ref{parameter2} lists hyper-parameters utilized for the customized network fine-tuning.
\begin{table}[t]
\centering
\renewcommand\arraystretch{1}
\caption{Hyper-Parameter for Network Fine-Tuning}
\label{parameter2}
\begin{tabular}{c|c}
\toprule 
\textbf{Parameter  }       & \textbf{Value} \\ \midrule
Batch size      &  32     \\ \midrule
Starting learning rate  &  $1\times 10 ^{-3}$      \\ \midrule
Learning rate scheduler &   Epochs 40 and 80    \\ \midrule
Learning-rate decaying factor &   0.5    \\ \midrule
Epochs &   100  \\ \midrule
Optimizer & ADAM      \\ \midrule
Loss function & MSELoss     \\ 
\bottomrule 
\end{tabular}
\end{table}
The training of the network is performed by PyTorch by utilizing the adaptive moment estimation (ADAM) optimizer~\cite{diederik2014adam}.

\subsubsection{Simulation Results and Analysis of Scatterer Recognition}
To verify the performance of the developed ScaR algorithm, two evaluation metrics are proposed.
The first one is the accuracy of the classification task of whether there are scatterers in grids. 
For a scatterer grid map with a dimension of $s_{\rm x}\times s_{\rm y}$, the classification accuracy can be expressed as 
\begin{equation}
    P_{\rm cla}=\frac{N_{\rm z}+N_{\rm nz}}{s_{\rm x}\times s_{\rm y}}
\end{equation}
where $N_{\rm z}$ represents the number of grids with zero labels and zero predicted values for scatterer density. $N_{\rm nz}$ represents the number of grids with non-zero labels and non-zero predicted values for scatterer density.
The second metric is the accuracy of scatterer density in grids containing scatterers, which is denoted by
\begin{equation}
    P_{\rm reg}=1-\frac{\sum |d_{\rm p}-d_{\rm g}|}{\sum d_{\rm g}}
\end{equation}
where $d_{\rm p}$ and $d_{\rm g}$ are the predicted value and ground truth of scatterer density in a grid, respectively.
$\sum$ represents the sum of all grids in grid maps participating in the test.

\begin{figure}[t]
\centering
\includegraphics[width=3in]{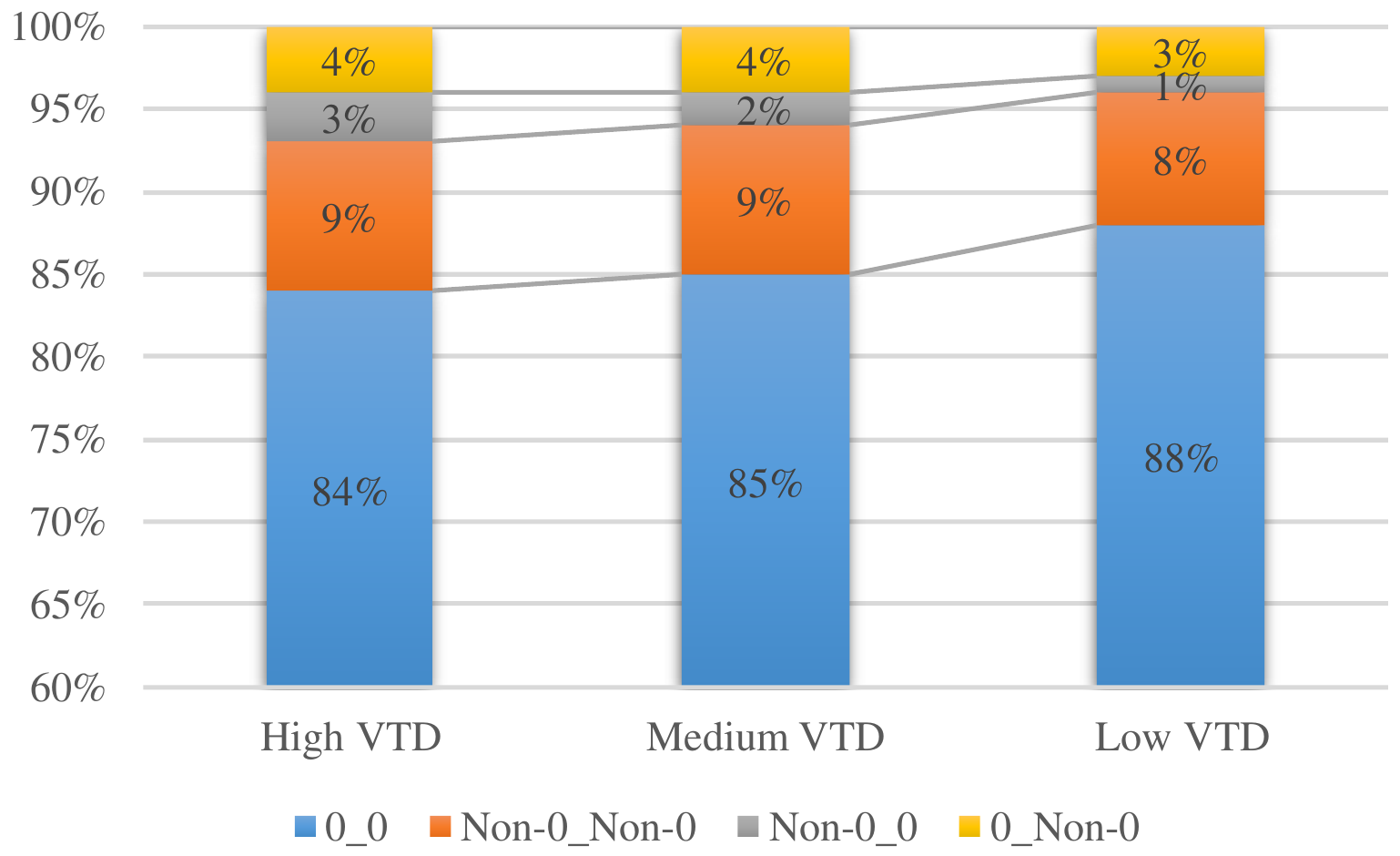}
\caption{Accuracy of scatterer grid classification under high, medium, and low VTDs.}
\label{cla}
\end{figure}

For metric one, Fig. \ref{cla} illustrates the accuracy of scatterer grid classification under high, medium, and low VTDs at the urban crossroad.
In Fig. \ref{cla}, ``$0\_0$", ``Non-0$\_$Non-0", ``Non-0$\_$0", and ``0$\_$Non-0" represents ``ground truth is 0 and predicted value is 0", ``ground truth is not 0 but predicted value is 0", ``ground truth is 0 but predicted value is not 0", and ``ground truth is not 0 and predicted value is not 0", respectively.
``$0\_0$" and ``Non-0$\_$Non-0" are cases where the classification results are correct.
As shown in Fig. \ref{cla}, the classification accuracy exceeds 93\% for all three VTDs. 
In addition, as the VTD decreases, the classification accuracy gradually increases. 
This is attributed to the reduction in the number of vehicles, leading to decreased complexity in physical and electromagnetic spaces.
Consequently, it becomes easier to extract complex nonlinear mapping relationships between physical and electromagnetic spaces.

\begin{figure}[t]
\centering
\includegraphics[width=3.5in]{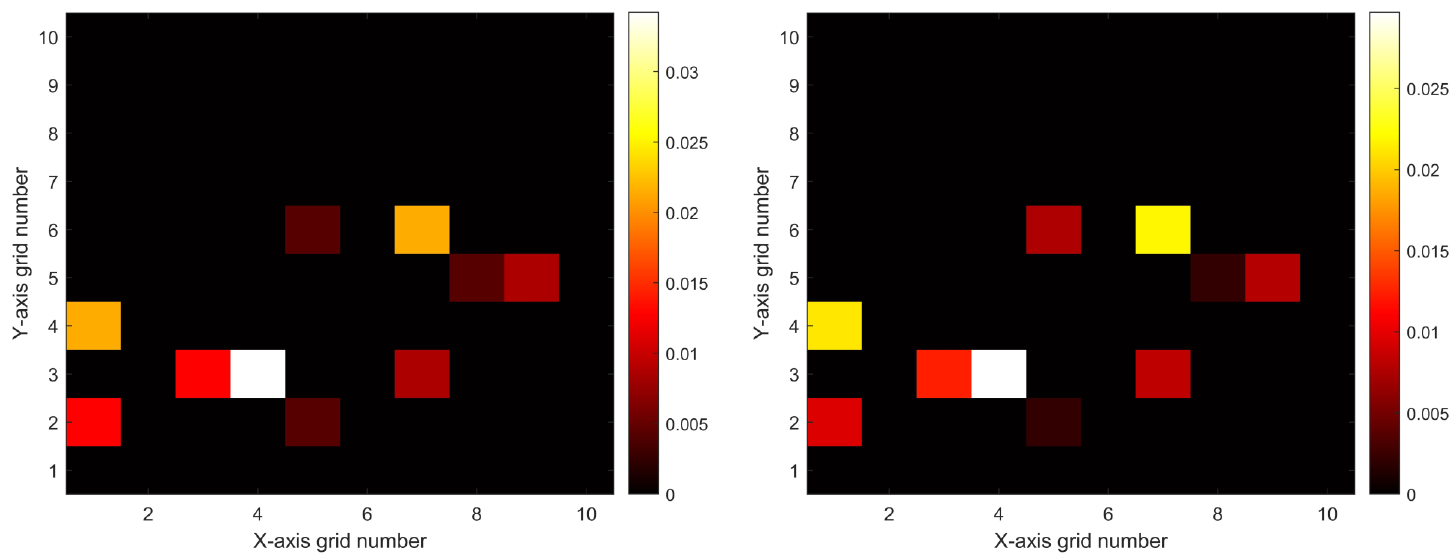}
\caption{Comparison between the predicted value and ground truth of the scatterer grid map.}
\label{scatterer-result}
\end{figure}

For the second metric, Table \ref{reg} presents the accuracy of scatterer density under high, medium, and low VTDs at the urban crossroad.
\begin{table}[t]
\centering
\renewcommand\arraystretch{1.1}
\caption{Accuracy of Scatterer Density Under High, Medium, and Low VTDs}
\label{reg}
\begin{tabular}{cc}
\hline
Scenario conditions & Accuracy of scatterer density \\ \hline
High VTD            & 85\% \\
Medium VTD          & 88\% \\
Low VTD             & 89\% \\ \hline
\end{tabular}
\end{table}
Similarly, as the VTD decreases, the accuracy of scatterer density gradually increases due to the reduced complexity in physical environment and electromagnetic space.
Fig. \ref{scatterer-result} presents a comparison between the predicted value and ground truth of the scatterer grid map in a snapshot under the medium VTD at the urban crossroad.
In Fig. \ref{scatterer-result}, it can be seen that there is solely one grid error about the position of the scatterer, and the predicted density of the scatterer is highly consistent with the ground truth.

\subsection{Model Simulation}
\begin{figure}[t]
\includegraphics[width=3.5in]{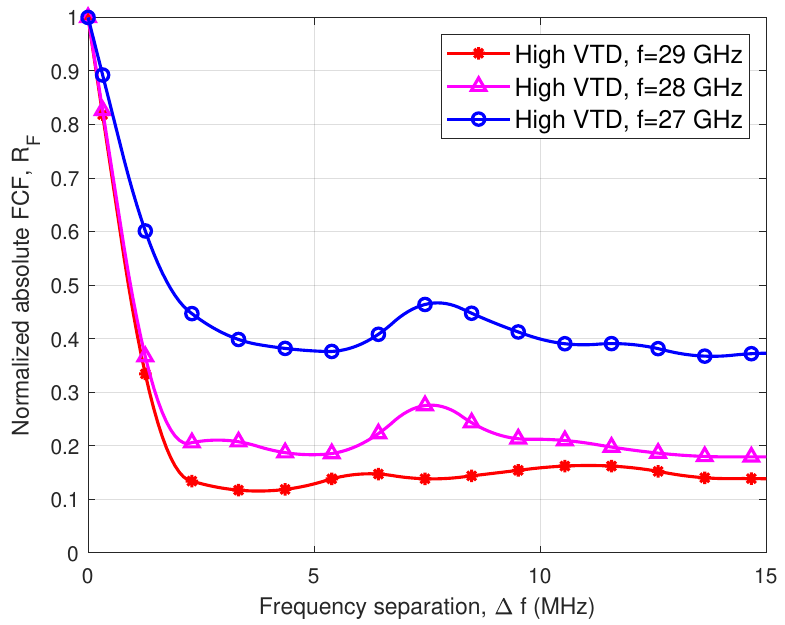}
\caption{Absolute normalized FCFs for different frequencies under the high VTD ($D_{\rm tcv}(t)$ = 60 m, $\mathbf{v}^{\rm T}(t)$ = 12 m/s, $\mathbf{v}^{\rm R}(t)$ = 15 m/s, $\overline{\mathbf{v}}_l^{\rm d,T}(t)$ = 10 m/s, $\overline{\mathbf{v}}_l^{\rm d,R}(t)$ = 10 m/s).}
\label{FCF-f}
\end{figure}

Fig. \ref{FCF-f} shows absolute normalized FCFs for different frequency bands under the high VTD at the urban crossroad.
In Fig. \ref{FCF-f}, comparing FCFs under different frequency bands, it can be observed that FCFs not only depend on frequency but also frequency separation, demonstrating the capturing of frequency non-stationary of the proposed channel model. 
In addition, compared to the lower frequency band of 27 GHz, FCFs of the higher frequency band are relatively lower. 
The philosophy is that the channel characteristics in the higher frequency band are more complex, resulting in lower frequency correlation.

\begin{figure}[t]
\includegraphics[width=3.5in]{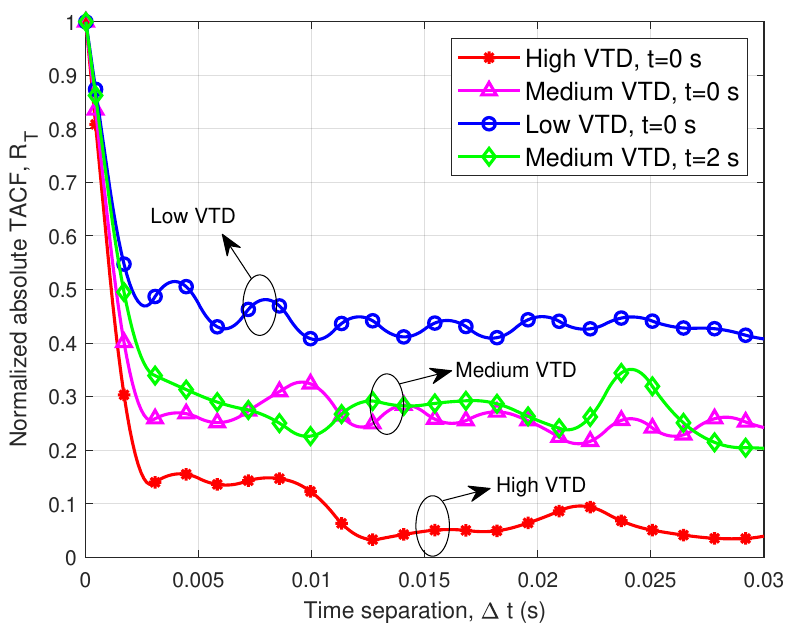}
\caption{Absolute normalized TACFs under high, medium, and low VTDs ($D_{\rm tcv}(t)$ = 80 m, $\mathbf{v}^{\rm T}(t)$ = 19 m/s, $\mathbf{v}^{\rm R}(t)$ = 17 m/s, $\overline{\mathbf{v}}_l^{\rm d,T}(t)$ = 12 m/s, $\overline{\mathbf{v}}_l^{\rm d,R}(t)$ = 14 m/s).}
\label{TACF-VTD}
\end{figure}

Fig. \ref{TACF-VTD} shows absolute normalized TACFs under high, medium, and low VTDs at the urban crossroad.
In Fig. \ref{TACF-VTD}, comparing TACFs at different time instants under the same VTD, it can be seen that TACFs not only depend on the time instant but also on the time separation, thus modeling time non-stationarity.
In addition, by comparing TACFs under different VTDs, it can be concluded that as the VTD decreases, TACFs significantly increase. 
This is because that, as the number of vehicles in the environment decreases, the channels are less complex and dynamic, resulting in higher temporal correlation.

\subsection{Model Verification}
The deterministic channel model based on the RT technology is of high accuracy.
Specifically, relying on the Wireless InSite simulation platform~\cite{InSite}, the RT technology is exploited to obtain the CIR in the scenario shown in Fig. \ref{WI_AirSim_scenario}.
Based on the CIR, the generality of the proposed channel model is verified.

\begin{figure}[t]
\includegraphics[width=3.5in]{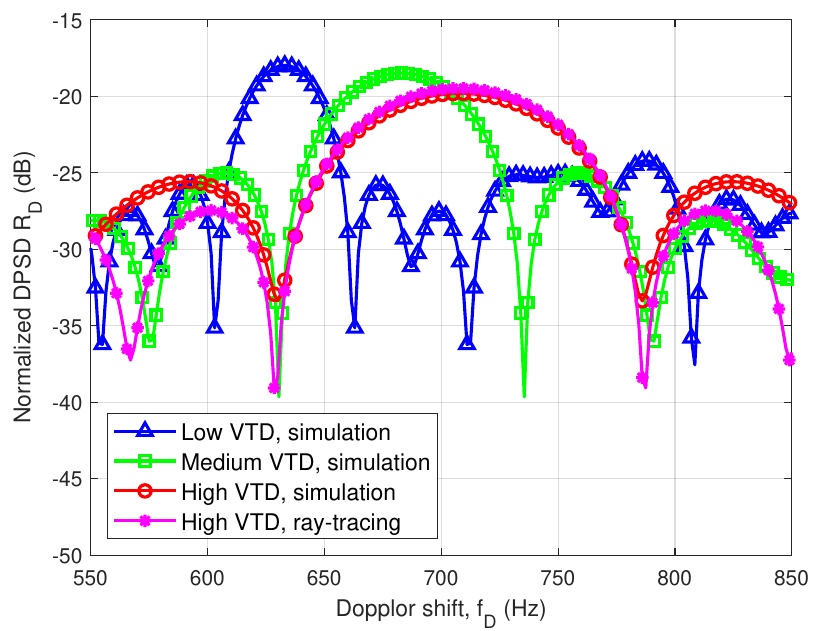}
\caption{Comparison of simulated DPSDs and RT-based DPSDs ($D_{\rm tcv}(t)$ = 80 m, $\mathbf{v}^{\rm T}(t)$ = 12 m/s, $\mathbf{v}^{\rm R}(t)$ = 8 m/s, $\overline{\mathbf{v}}_l^{\rm d,T}(t)$ = 7 m/s, $\overline{\mathbf{v}}_l^{\rm d,R}(t)$ = 5 m/s).}
\label{DPSDs-comparison}
\end{figure}

Through (\ref{TF-CF}) and (\ref{DPSD}), the RT-based CIR under the medium VTD is processed to obtain the RT-based DPSD. 
Additionally, the simulated DPSDs under high, medium, and low VTDs based on the proposed channel model are solved.
The aforementioned RT-based DPSDs and simulated DPSDs are shown in Fig. \ref{DPSDs-comparison}.
In Fig. \ref{DPSDs-comparison}, the simulated DPSDs are compared with the RT-based DPSDs under high VTD. 
It can be observed that simulation results can match well with RT-based results, which confirms the generality of the proposed model.
In addition, as the VTD increases, the distributions of DPSDs become flatter. 
This phenomenon is explained that, at higher VTDs, the received power tends to come from vehicles around Tx and Rx over more directions.

\begin{figure}[t]
\includegraphics[width=3.5in]{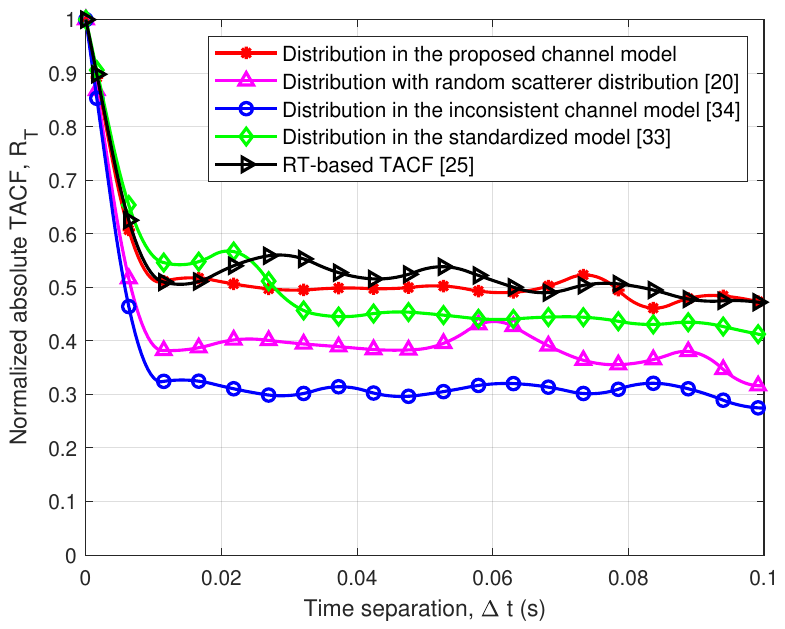}
\caption{Comparison of simulated TACFs and RT-based TACFs ($D_{\rm tcv}(t)$ = 60 m, $\mathbf{v}^{\rm T}(t)$ = 14 m/s, $\mathbf{v}^{\rm R}(t)$ = 11 m/s, $\overline{\mathbf{v}}_l^{\rm d,T}(t)$ = 10 m/s, $\overline{\mathbf{v}}_l^{\rm d,R}(t)$ = 10 m/s).}
\label{TACF-comparison}
\end{figure}

Through (\ref{TF-CF}), the RT-based CIR under low VTD is processed to obtain the RT-based TACF. 
In Fig. \ref{TACF-comparison}, the RT-based TACF~\cite{InSite} is compared with the TACF obtained from standardized models~\cite{generation2019technical}, TACF generated based on the random statistical distribution of scatterer positions~\cite{huang2024lidar}, TACF based on the channel model without considering consistency~\cite{huang20213}, and TACF obtained from the proposed model.
In Fig. \ref{TACF-comparison}, it can be seen that  the TACF obtained by the proposed model that captures time consistency can achieve the close fit with the RT-based TACF. 
For the TACF generated based on standardized models and the TACF obtained from random scatterer positions, due to the lack of accurate scatterer positions and quantities, the results cannot be closely fitted with those based on RT. 
For the TACF under time inconsistency, the smooth evolution of the channel in the time domain is ignored, which cannot capture the correlation between channels at adjacent moments.
Therefore, the TACF under time inconsistency is lower and cannot fit well with the RT-based TACF.

\section{Conclusions}
This paper has proposed a novel multi-modal intelligent V2V channel model from LiDAR point clouds based on SoM.
Based on a new V2V mixed sensing-communication integration simulation dataset, a novel ScaR algorithm has been developed to recognize scatterers from LiDAR point clouds.
In the ScaR algorithm, the customized network, which has built on the SegNet architecture, has been utilized to explore the complex and nonlinear mapping relationship between physical environment and electromagnetic space.
Simulation results have demonstrated that the classification accuracy of scatterers exceeds 93\% and the accuracy of scatterer density exceeds 85\% under high, medium, and low VTDs.
By utilizing LiDAR point clouds to recognize scatterers and distinguishing dynamic and static scatterers, channel non-stationarity and consistency have been captured.
Based on the proposed model, important channel statistical properties have been derived and analyzed.
Compared to the low VTD, the proposed V2V channel model has exhibited a lower temporal correlation and has possessed a flatter distribution of DPSD under the high and medium VTDs.
The close fit between simulation results and RT-based results has demonstrated the necessities of the mapping relationship exploration and the accuracy of the proposed model.
\IEEEpubidadjcol

\newpage

\vfill

\end{document}